\documentclass[aps,prb,amsmath,amssymb,twocolumn,10pt,superscriptaddress]{revtex4-2}

\usepackage{times,txfonts}
\usepackage{gensymb}
\usepackage{siunitx}
\usepackage{dcolumn}
\usepackage{multirow}
\usepackage[normalem]{ulem}
\usepackage{times}
\usepackage[utf8]{inputenc}
\DeclareUnicodeCharacter{2212}{-}
\usepackage[T1]{fontenc}
\usepackage[]{graphicx}
\usepackage[dvipsnames]{xcolor}
\usepackage{tabularx}
\usepackage{float}
\usepackage{amsfonts}
\usepackage{amsmath}
\usepackage{amssymb}
\usepackage{bm}
\usepackage{enumerate}
\usepackage{mathtools}
\usepackage[breaklinks,colorlinks=true,citecolor=blue]{hyperref}

\begin{document}

\title{Spin-orbit entangled moments and magnetic exchange interactions in cobalt-based honeycomb magnets BaCo$_2$($X$O$_4$)$_2$ ($X$ = P, As, Sb)}

\author{Subhasis Samanta}
\email{samanta@skku.edu}
\thanks{These authors contributed equally.}
\affiliation{Department of Semiconductor Physics and Institute of Quantum Convergence Technology, Kangwon National University, Chuncheon 24341, Republic of Korea}
\affiliation{Center for Extreme Quantum Matter and Functionality, Sungkyunkwan University, Suwon 16419, Republic of Korea}

\author{Fabrizio Cossu}
\thanks{These authors contributed equally.}
\affiliation{Department of Semiconductor Physics and Institute of Quantum Convergence Technology, Kangwon National University, Chuncheon 24341, Republic of Korea}
\affiliation{Department of Physics, School of Natural and Computing Sciences, University of Aberdeen, Aberdeen, AB24 3UE, United Kingdom}
\affiliation{School of Physics, Engineering and Technology, University of York, Heslington, York YO10 5DD, United Kingdom}

\author{Heung-Sik Kim}
\email{heungsikim@kangwon.ac.kr}
\affiliation{Department of Semiconductor Physics and Institute of Quantum Convergence Technology, Kangwon National University, Chuncheon 24341, Republic of Korea}

\begin{abstract}
Co-based honeycomb magnets have been actively studied recently for the potential realization of emergent quantum magnetism therein such as the Kitaev spin liquid. Here we employ density functional and dynamical mean-field theory methods to examine a family of the Kitaev magnet candidates BaCo$_2$($X$O$_4$)$_2$ ($X$ = P, As, Sb), where the compound with $X$ = Sb being not synthesized yet. Our study confirms the formation of Mott insulating phase and the $J_{\rm eff}$ = 1/2 spin moments at Co$^{2+}$ sites despite the presence of a sizable amount of trigonal crystal field in all three compounds. The pnictogen substitution from phosphorus to antimony significantly changes the in-plane lattice parameters and direct overlap integral between the neighboring Co ions, leading to the suppression of the Heisenberg interaction. More interestingly, the marginal antiferromagnetic nearest-neighbor Kitaev term changes sign into a ferromagnetic one and becomes sizable at the $X$ = Sb limit. Our study suggests that the pnictogen substitution can be a viable route to continuously tune magnetic exchange interactions and to promote magnetic frustration for the realization of potential spin liquid phases in BaCo$_2$($X$O$_4$)$_2$. 
\end{abstract}

\maketitle

\section*{Introduction}

The study of Kitaev spin liquid (KSL) \cite{Savary2016,Ng2017,Nagler2019,Hickey2022,Young2022,Park2022} has recently captured much attention because of its potential use in fault-tolerant topological quantum computation \cite{Kitaev2006,Kitaev2003,Sarma2008}. For material realization, early studies suggested 4$d$- and 5$d$-transition metal compounds with strong spin-orbit coupling as potential candidates \cite{Khaliullin2009,Khaliullin2010,Young2014,Kim2015,Kee2014}. However, subsequent studies showed that these materials host long-range magnetic order at low temperatures due to the appreciable presence of other interactions such as third-nearest-neighbor Heisenberg and non-cubic distortion \cite{Valenti2016}. While the search for more ideal candidates for KSL was ongoing, two recent theoretical works concomitantly proposed that cobalt-based compounds with the high-spin $d^7$ configuration can host the formation of pseudospin-1/2 magnetic moments, which is an essential component for the realization of the Kitaev exchange interactions \cite{Khaliullin2018,Motome2018}.

Among the materials candidates, BaCo$_2$(AsO$_4$)$_2$ (BCAO) \cite{Regnault1977,Regnault2018,Cava2020,Wang2021,Armitage2023,Li2023,Winter2022,Kim2022,Paramekanti2021,Broholm2023,Paramekanti2023,Streltsov2022} and BaCo$_2$(PO$_4$)$_2$ (BCPO) \cite{Harrison1998,Ross2018,Cava2018,Lorenz2023} have attracted much attention recently. It has been reported that exerting magnetic fields in the honeycomb plane can induce two successive magnetic transitions in BCAO and BCPO at low temperatures, analogous to the well-known behavior observed \cite{Regnault2018,Cava2018,Cava2020,Wang2021,Armitage2023,Lorenz2023} in 4$d$-based $\alpha$-RuCl$_3$, where the external magnetic field is considered to suppress the long-range order and to reveal the underlying KSL phase \cite{Matsuda2018,Matsuda2021,Takagi2022}. On the other hand, $ab$ $initio$ and model Hamiltonian studies suggest that the strength of the Kitaev interactions in BCPO and BCAO is weak due to the dominant direct overlap integrals between nearest-neighboring cobalt $d$-orbitals, and that nearest-neighbor and third-nearest-neighbor Heisenberg exchange interactions are the most dominant contributions in these systems \cite{Winter2022,Kim2022,Paramekanti2021,Broholm2023,Paramekanti2023}.

\begin{figure*}
\begin{center}
\includegraphics[angle=-0,origin=c,scale=0.355]{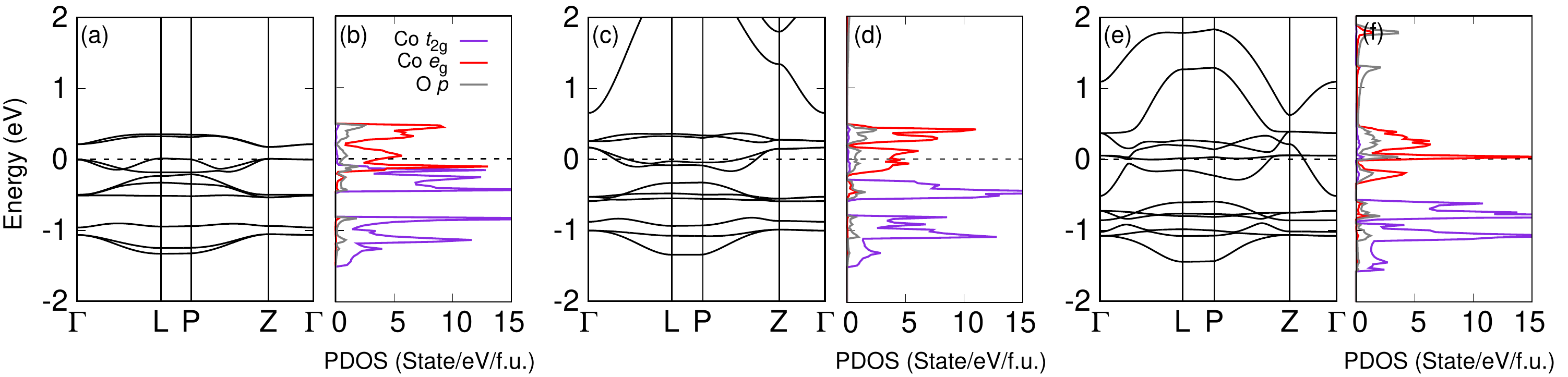}
\caption{{\bf Electronic band structure and PDOS.} Non-spin polarized band structure alongside PDOS (Co-$t_{\rm 2g}$ (violet), -$e_{\rm g}$ (red), and O-$p$ (gray)) obtained from LDA calculations for {\bf a, b} BCPO, {\bf c, d} BCAO, and {\bf e, f} BCSO.}
\label{fig:Fig1}
\end{center}
\end{figure*}

Inspired by recent experimental and theoretical developments on BCPO and BCAO, here we investigate a whole series of three cobaltate compounds with the chemical formula BaCo$_2(X$O$_4$)$_2$ (BC$X$O) ($X$ = P, As, Sb), in which BCSO has not been synthesized yet. The major motivation of the pnictogen substitution is the following: as we replace $X$ with P, As, and Sb, the in-plane lattice constants of BC$X$O should systematically increase, which affect the direct $d$-$d$ overlap integrals of Co and, ultimately suppress the strength of nearest-neighbor Heisenberg exchange interactions. On the other hand, the pnictogen substitution can tune the size of the trigonal distortion of CoO$_6$ octahedra. The smaller trigonal distortion is always good for KSL since atomic spin-orbit coupling strength of Co is already small compared to 4$d$ and 5$d$ transition metals. Therefore, the presence of sizable trigonal distortion could lead to the quenching of $L_{\rm eff}=1$ orbital degree of freedom, detrimental for realization of the spin-orbital-entangled magnetic moments and spin liquid phases.

For the study on the nature of magnetic local moments and exchange interactions of BC$X$O series, in this work we employ first-principles density functional plus dynamical mean-field theory (DFT+DMFT) and Wannierization techniques in combination with recently developed perturbative expressions of the exchange interactions for $d^7$ cobaltates \cite{Kee2023}. Our DFT+DMFT calculations confirm the formation of unquenched $L_{\rm eff}=1$ with $S=3/2$ spins at Co$^{2+}$ ions in all three compounds, which forms the $J_{\rm eff}=1/2$ in the presence of spin-orbit coupling. From the comparison of trigonal crystal field-splitting, we find that the size of trigonal crystal fields within the $t_{\rm 2g}$ orbital gradually reduces and eventually changes sign as the pnictogen element becomes heavier (P$\rightarrow$As$\rightarrow$Sb), showing that pnictogen substitution can be a way to tune the crystal fields in this system. Furthermore, the estimation of magnetic exchange interactions based on our first-principles calculations reveals that {\it i}) the suppression of the $d$-$d$ direct overlap between the neighboring Co ions leads to weaker Heisenberg interactions and {\it ii}) the sign change of the Kitaev interaction from the antiferromagnetic to ferromagnetic in the nearest-neighbor bondings as we approach from the $X$ = P to Sb limits. In addition, the ratio between the nearest and third-nearest Heisenberg interactions can be also chemically tuned, which can be utilized for the achievement of the geometric frustration via the competition between the nearest- and third-nearest-neighbor Heisenberg interactions in this system \cite{Broholm2023,Reatto1979,Fouet2001}. Overall, our result shows that the BC$X$O family is an interesting system with chemically tunable exchange interactions, where geometric and Kitaev-type exchange frustrations can coexist to realize novel magnetism.

\begin{figure*}
\begin{center}
\includegraphics[angle=-0,origin=c,scale=0.34]{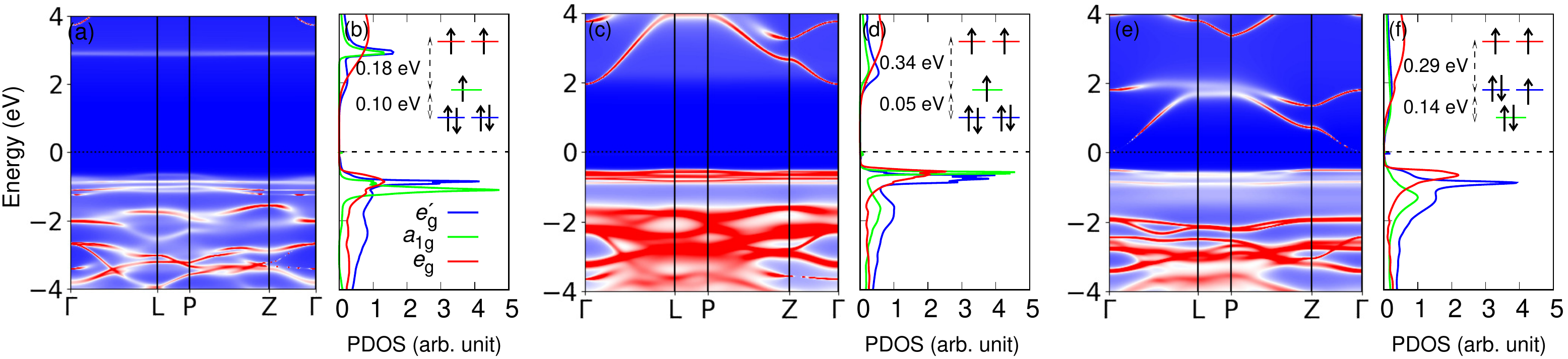}
\caption{{\bf Spectral functions and PDOS from DMFT.} Momentum- and frequency-dependent spectral functional alongside momentum integrated projected density of states, plotted for {\bf a, b} BCPO, {\bf c, d} BCAO, and {\bf e, f} BCSO. The results are obtained employing Ising-type Coulomb interactions at 116 K. The insets in the PDOS plots show the crystal field splittings within the Co $d$-orbitals.}
\label{fig:Fig2}
\end{center}
\end{figure*}

\section*{Results}

\subsection*{Electronic structure from nonmagnetic DFT}

The crystal structures of BCAO and BCPO compounds possess rhombohedral geometry with the $R\bar{3}$ space group symmetry \cite{Tamara2008,Harrison1998}. We assume that BCSO, which is a fictitious compound with no experimental report yet, has the same $R\bar{3}$ structure as BCPO and BCAO. Optimizations of structural parameters show that the substitution of the pnictogen element from P to Sb leads to the enhancement of the in-plane lattice parameters (see Table \ref{tab:Tab1}) and the distance between nearest-neighboring Co sites. On the other hand, systematic enhancement of the $X$O$_4$ tetrahedral volume as a function of the pnictogen substitution (from P to As to Sb) results in contraction of CoO$_6$ octahedral volume, which should directly impact both intra- and inter-site hoppings. Note that such tendencies mentioned above can be checked in the experimental structures of BCPO and BCAO as well, despite the lattice parameters and Co-O bond lengths being slightly underestimated in our calculation results (see Table \ref{tab:Tab1}). 

\begin{table}
\caption{Lattice constants of BC$X$O obtained from DFT+$U$ calculations and reported one in the literature.}
\centering
\begin{tabular}{c|l|l}
\hline\hline
Compound & \multicolumn{2}{c}{Lattice constants and bond lengths} \\
\cline{2-3}
 & \multicolumn{1}{c|}{DFT+$U$ optimized} & \multicolumn{1}{c}{Experiment} \\
\hline
 & $a=$ 4.833 \AA, $c=$ 23.070 \AA & $a=$ 4.855 \AA, $c=$ 23.215 \AA \\
BCPO & $\alpha=90^{\circ}$, $\gamma=120^{\circ}$ & $\alpha=90^{\circ}$, $\gamma=120^{\circ}$ \\
 & Co1$-$O2 = 2.062 \AA & Co1$-$O2 = 2.083 \AA \\
 & Co1$-$O2 = 2.089 \AA & Co1$-$O2 = 2.112 {\AA} \cite{Harrison1998} \\
 & $\angle$Co-O-Co = 84.622$^{\circ}$ & \\
\hline
 & $a=$ 4.999 \AA, $c=$ 23.227 \AA & $a=$5.007 \AA, $c=$ 23.491 \AA \\
BCAO & $\alpha=90^{\circ}$, $\gamma=120^{\circ}$ & $\alpha=90^{\circ}$, $\gamma=120^{\circ}$ \\
 & Co1$-$O2 = 2.056\AA & Co1$-$O2 = 2.079 \AA \\
 & Co1$-$O2 = 2.077 & Co1$-$O2 = 2.101 {\AA} \cite{Tamara2008} \\
 & $\angle$Co-O-Co = 88.738$^{\circ}$ & \\
\hline
 & $a=$ 5.179 \AA, $c=$ 23.414 \AA & \\
BCSO & $\alpha=90^{\circ}$, $\gamma=120^{\circ}$ & \\
 & Co1$-$O2 = 2.054 \AA & \hspace*{1.5cm} - \\
 & Co1$-$O2 = 2.063 \AA & \\
 & $\angle$Co-O-Co = 93.288$^{\circ}$ & \\
\hline\hline
\end{tabular}
\label{tab:Tab1}
\end{table}

Figure \ref{fig:Fig1} plots nonmagnetic band structures and projected densities of states (PDOS) for BCPO, BCAO, and BCSO without including on-site Coulomb repulsion and spin-orbit coupling. The lower-lying six fully occupied bands are derived from Co-$t_{\rm 2g}$ orbitals while partially occupied Co-$e_{\rm g}$ bands are located around $E_{\rm F}\pm$0.3 eV. Inspecting the almost flat out-of-plane band dispersion of the Co $t_{\rm 2g}$- and $e_{\rm g}$-bands along $Z-\Gamma$, we find that all BC$X$O compounds host almost perfect two-dimensional nature. Note that in BCSO, the presence of Sb $s$-orbital degree of freedom introduces dispersive bands along the $Z-\Gamma$ direction at the Fermi level (see Fig. \ref{fig:Fig2}e), which is pushed above the Fermi level as the Coulomb repulsion is introduced (see below). The PDOS clearly shows crystal field-splitting between the $t_{\rm 2g}$ and $e_{\rm g}$ states.

\subsection*{Evidence of formation of $L_{\rm eff}=1$ from DMFT}

\begin{figure}
\centering
\includegraphics[angle=-0,origin=c,scale=0.34]{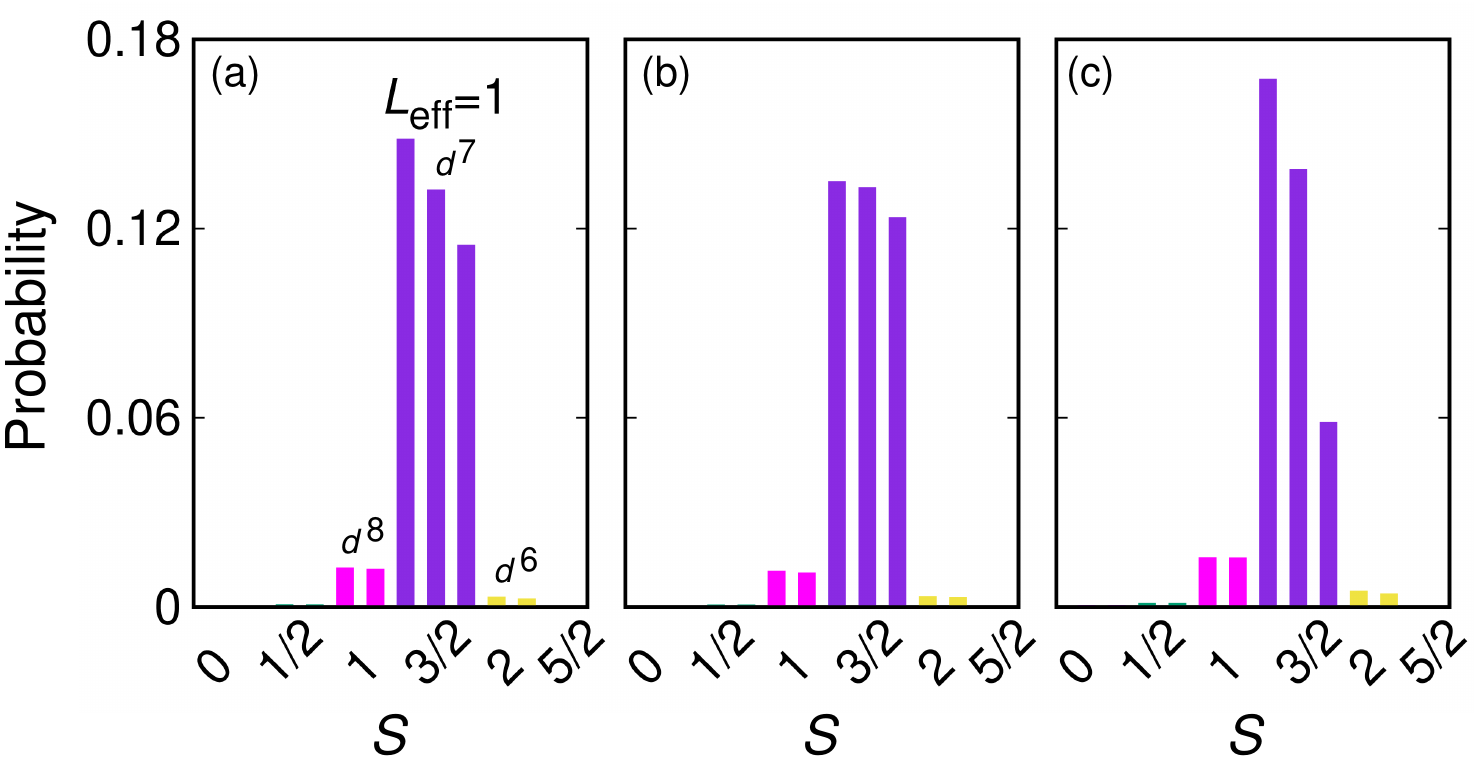}
\caption{{\bf Histogram of Co $d$ atomic configurations.} {\bf a-c} Probability of atomic multiplet states plotted for BCPO, BCAO, and BCSO, respectively with different spin configurations showing the formation active $L_{\rm eff}=1$ orbital degree of freedom.}
\label{fig:Fig3}
\end{figure}

\begin{table*}[]
\centering
\caption{First and third nearest-neighbor $d$-$d$ hopping parameters, $p$-$d$ hopping integrals, and cubic crystal-field splitting in meV for BCPO, BCAO, and BCSO, obtained from nonmagnetic DFT calculations. Noteworthy to mention that hopping parameters are listed for one of the three-fold symmetric Co-Co bonds. The same can be obtained for other two bonds through cyclic rotation.}
\setlength{\tabcolsep}{0.26em}
\begin{tabular}{c | r r r r r r | r r r r r r | r r | c}
\hline\hline
Compound & \multicolumn{6}{c|}{First nearest-neighbor hopping} & \multicolumn{6}{c|}{Third nearest-neighbor hopping} & \multicolumn{2}{c|}{$p$-$d$ hopping} & \multicolumn{1}{c}{Crystal-field} \\
\cline{2-16}
 & \multicolumn{1}{c}{$t_1$} & \multicolumn{1}{c}{$t_2$} & \multicolumn{1}{c}{$t_3$} & \multicolumn{1}{c}{$t_4$} & \multicolumn{1}{c}{$t_5$} & \multicolumn{1}{c|}{$t_6$} & \multicolumn{1}{c}{$t'_1$} & \multicolumn{1}{c}{$t'_2$} & \multicolumn{1}{c}{$t'_3$} & \multicolumn{1}{c}{$t'_4$} & \multicolumn{1}{c}{$t'_5$} & \multicolumn{1}{c|}{$t'_6$} & \multicolumn{1}{c}{$t_{pd\sigma}$} & \multicolumn{1}{c|}{$t_{pd\pi}$} & \multicolumn{1}{c}{$\Delta$} \\
\hline
BCPO & 75.33 & $−$33.44 & $−$381.57 & $−$47.31 & $−$97.13 &  14.23 & 1.11 & $−$11.90 & $−$35.66 & $−$33.06 & 120.22 & 12.89 & 1338.00 & 615.86 &  850.18 \\
BCAO & 71.49 & $−$20.63 & $−$294.43 & $−$43.73 & $−$36.15 &  66.04 & 0.83 &  $−$2.53 & $−$48.31 & $−$40.92 & 135.34 & 26.18 & 1353.12 & 644.50 &  894.34 \\
BCSO & 39.87 & 34.64 & $−$155.84 & 71.82 & 5.96 & 120.26 & 1.88 & 4.35 & $−$34.11 & $−$22.10 &  58.88 & 12.05 & 1209.37 & 566.21 & 1072.45 \\
\hline\hline
\end{tabular}
\label{tab:Tab2}
\end{table*}

To gain more insight into the formation of paramagnetic Mott insulating phase, the presence of active $L_{\rm eff}=1$ orbital degree of freedom with $S=3/2$ spins of Co$^{2+}$ ions, and for the estimation of the size of trigonal crystal field splitting, we carried out DMFT calculations for BC$X$O. The accurate crystal structures were obtained by further invoking atomic relaxation within DMFT calculations. We mention here that simple DFT or DFT+$U$ calculations fail to capture paramagnetic Mott insulating features and the presence of unquenched active orbital degrees of freedom in the high-spin $d^7$ configuration \cite{Kim2022}.

Figure \ref{fig:Fig2} shows the LDA+DMFT spectral functions of BC$X$O. The results are obtained using the Ising-type Coulomb interaction for ($U, J_{\rm H}$) = (8, 0.8) eV at 116 K. The key feature of the spectra is the formation of a clear Mott gap in all three compounds. From the spectra, it is quite evident that the size of the Mott gap gradually decreases, where dispersive conduction bands with strong pnictogen character move closer to the Fermi level as we go from P to Sb. Panels of Figs. \ref{fig:Fig2}b, d, and f show PDOS of BCPO, BCAO, and BCSO, respectively. Crystal field levels of the $a_{\rm 1g}$, $e^\prime_{\rm g}$, and $e_{\rm g}$ orbitals can be obtained as self-energies in the infinite frequency limit from the quantum Monte Carlo impurity solver, which are depicted in the insets in each of the PDOS panels. We find that {\it i}) the size of the $t_{\rm 2g}$-$e_{\rm g}$ splitting is increased as we go from $X$ = P to Sb, consistently with the concomitant volume contraction of the CoO$_6$ octahedron. {\it ii}) The size of trigonal crystal field decreases and eventually reverses sign as the pnictogen element changes from P to Sb. The splitting between the $a_{\rm 1g}$ and $e^\prime_{\rm g}$ becomes smallest at $X$ = As, which may be suppressed to zero when a small amount of Sb replacement of As in BCAO is introduced. Importantly, in all three compounds, the size of trigonal splitting is smaller than or at largest comparable to the size of spin-orbit coupling within the Co $d$-orbitals.

The Monte Carlo probability of atomic multiplets within the Co $d$-orbitals for all three BC$X$O compounds are shown in Fig. \ref{fig:Fig3}. Here we are focusing on paramagnetic states. So we show probabilities of one spin component only (probabilities of their Kramers partners are the same because of the time-reversal symmetry). Firstly, the DMFT results show that Co $d$-orbitals are mostly in the $d^7$ high-spin ($S=3/2$) sector, confirming the strong suppression of charge fluctuations and the presence of the Mott-insulating character across the whole BC$X$O series. Secondly, the presence of three major $S=3/2$ $d^7$ multiplet states can be seen, confirming the presence of the unquenched $L_{\rm eff}=1$ orbital degree of freedom in all three compounds despite the presence of finite trigonal crystal fields. Hence, the inclusion of the atomic spin-orbit coupling should give rise to the formation of the $J_{\rm eff}=1/2$ doublet states in all three BC$X$O compounds, as suggested in BCAO previously \cite{Kim2022}. Furthermore, comparing $L_{\rm eff}=1$ states for BC$X$O, we find that the $L_{\rm eff}=1$ multiplet states in BCPO and BCAO are more equally populated than in BCSO. This can be attributed to the larger size of the trigonal crystal field in BCSO than those in other two compounds (compare insets of Fig. \ref{fig:Fig2}b, d, and f).

\subsection*{Hopping integrals and magnetic exchange interactions}

\begin{table*}[]
\centering
\caption{First and third nearest-neighbor exchange interactions of BCPO, BCAO, and BCSO in meV. Here, labels $A$ ($A'$), $B$ ($B'$), and $C$ ($C'$) indicate $t_{\rm 2g}$-$t_{\rm 2g}$, $t_{\rm 2g}$-$e_{\rm g}$, and $e_{\rm g}$-$e_{\rm g}$ exchange process between first (third) nearest-neighbors. The numbers 1, 2, and 3 denote direct, double-hole, and cyclic processes. The on-site Coulomb and Hund's coupling parameters were set to ($U, J_{\rm H})=$ (6, 1.2) eV for $d$ orbitals and ($U_{\rm p}, J_{\rm Hp})=$ (4, 1) eV for $p$, respectively.}
\setlength{\tabcolsep}{0.58em}
\begin{tabular}{c | c r r r r r r r r r r | c r r r r}
\hline\hline
Compound & \multicolumn{11}{c|}{First nearest-neighbor exchange} & \multicolumn{5}{c}{Third nearest-neighbor exchange} \\
\cline{2-17}
& & \multicolumn{1}{c}{$A1$} & \multicolumn{1}{c}{$B1$} & \multicolumn{1}{c}{$C1$} & \multicolumn{1}{c}{$A2$} & \multicolumn{1}{c}{$B2$} & \multicolumn{1}{c}{$C2$} & \multicolumn{1}{c}{$A3$} & \multicolumn{1}{c}{$B3$} & \multicolumn{1}{c}{$C3$} & \multicolumn{1}{c|}{Total} & & \multicolumn{1}{c}{$A'1$} & \multicolumn{1}{c}{$B'1$} & \multicolumn{1}{c}{$C'1$} & \multicolumn{1}{c}{Total}\\
\hline
& $J_1$ & $-$9.13 & 0 & 1.71 & 0.09 & 6.43 & $-$2.80 & 0.03 & $-$4.73 & 0 & $-$8.39 & $J_3$ & -0.06 & 0 & 2.28 & 2.22 \\
BCPO & $K_1$ & 3.52 & $-$0.01 & 0 & 0.22 & $-$3.62 & 0 & $-$0.31 & 2.36 & 0 & 2.16 & $K_3$ & $-$0.01 & $-$0.01 & 0 & $-$0.02 \\
& $\Gamma_1$ & $-$1.24 & 0 & 0 & 0 & 0 & 0  & 0 & 0 & 0 & $-$1.24 & $\Gamma_3$ & $-$0.04 & 0 & 0 & $-$0.04 \\
\hline
& $J_1$ & $-$5.97 & 0.08 & 0.47 & 0.11 & 8.17 & $-$3.37 & 0.04 & $-$6.10 & 0 & $-$6.56 & $J_3$ & $-$0.09 & 0.01 & 2.93 & 2.86 \\
BCAO & $K_1$ & 2.66 & $-$0.40 & 0 & 0.32 & $-$4.61 & 0 & $-$0.45 & 3.05 & 0 & 0.56 & $K_3$ & 0.01 & $-$0.06 & 0 & $-$0.05 \\
& $\Gamma_1$ & $-$0.59 & 0 & 0 & 0 & 0 & 0 & 0 & 0 & 0 & $-$0.59 & $\Gamma_3$ & $-$0.01 & 0 & 0 & $-$0.01 \\
\hline
& $J_1$ & $-$1.80 & 0.40 & 0.76 & 0.09 & 7.93 & $-$3.51 & 0.05 & $-$6.31 & 0 & $-$2.38 & $J_3$ & $-$0.05 & 0 & 0.58 & 0.53 \\
BCSO & $K_1$ & 0.68 & $-$1.32 & 0 & 0.35 & $-$4.58 & 0 & $-$0.50 & 3.15 & 0 & $-$2.21 & $K_3$ & 0.01 & $-$0.01 & 0 & 0 \\
& $\Gamma_1$ & 0.52 & 0 & 0 &  0 & 0 & 0 & 0 & 0 & 0 & 0.52 & $\Gamma_3$ & 0.01 & 0 & 0 & 0.01 \\
\hline\hline
\end{tabular}
\label{tab:Tab3}
\end{table*}

\begin{figure}
\centering
\includegraphics[angle=-0,origin=c,scale=0.22]{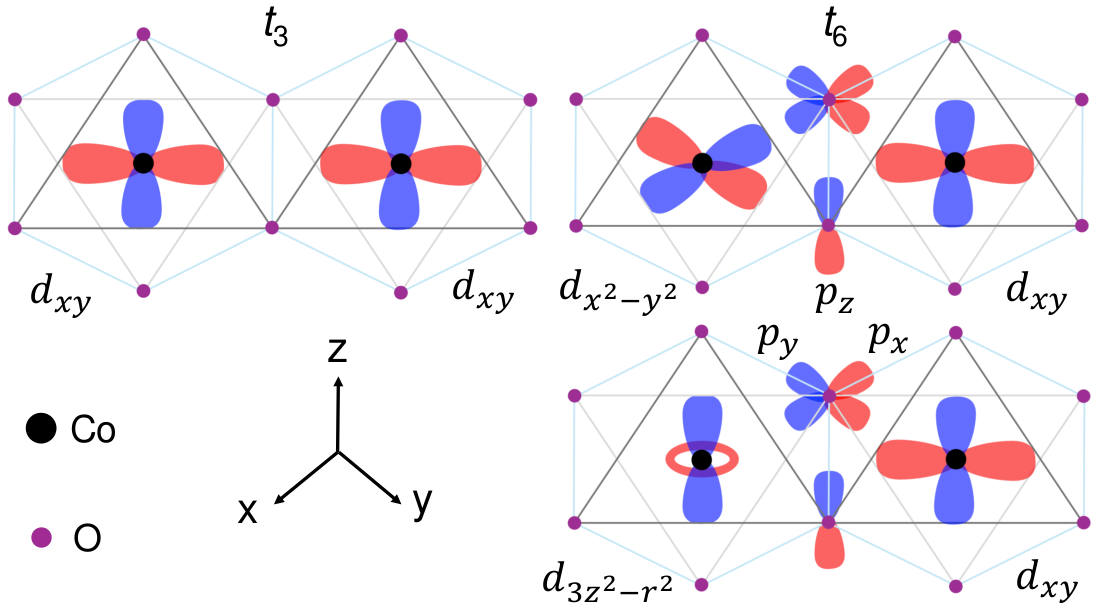}
\caption{{\bf Schematic representation of hopping channels.} Schematic diagram of two major hopping channels, depicting $d$-$d$ and $p$-$d$ hopping processes in BC$X$O compounds.}
\label{fig:Fig4}
\end{figure}

As the presence of the unquenched $L_{\rm eff}=1$ effective orbital degree of freedom which results in the formation of $J_{\rm eff}=1/2$ pseudospin in BC$X$O is checked, the following steps are adopted to estimate magnetic exchange interactions. First, hopping integrals between $d$-$d$ and $p$-$d$ hopping processes were computed from {\it ab initio} calculations. Then, we employed a recently developed strong-coupling perturbative expression for magnetic exchange interactions for $d^7$ cobaltates \cite{Kee2023}. Notably, this spin model has been shown to successfully explain the origin of weak Kitaev and dominant Heisenberg exchange interactions in BCAO, and may be applied for other types of honeycomb cobaltates as well \cite{Changhee2024}. For ideal edge-sharing octahedra on a honeycomb lattice with three-fold symmetric first and third nearest-neighbor interactions, Hamiltonian of the spin model can be written as
\begin{align}
H_{\rm spin} & = \sum_{\langle ij \rangle \in \gamma(\alpha\beta)}
\left[ J_1{\bf S}_i\cdot{\bf S}_j + K_1{S_i^\gamma}{S_j^\gamma} + \Gamma_1({S_i^\alpha}{S_j^\beta}+{S_i^\beta}{S_j^\alpha}) \right] \nonumber \\
& + \sum_{\langle\langle\langle ij \rangle\rangle\rangle \in \gamma(\alpha\beta)} \left[ J_3{\bf S}_i\cdot{\bf S}_j + K_3{S_i^\gamma}{S_j^\gamma} + \Gamma_3({S_i^\alpha}{S_j^\beta}+{S_i^\beta}{S_j^\alpha}) \right] \nonumber
\end{align}

Here, $\alpha, \beta, \gamma \in \{ x, y, z \}$ simultaneously represent the local octahedral coordinates for the directions of the spin $S=3/2$ and $L_{\rm eff}=1$ momenta and the corresponding bond directions. The $J_1$ ($J_3$), $K_1$ ($K_3$), and $\Gamma_1$ ($\Gamma_3$) stand for first- (third-) nearest-neighbor Heisenberg, Kitaev, and off-diagonal exchange terms, respectively. It must be noted that second-nearest-neighbor interactions are present in cobaltates, but they are much smaller in magnitude than first and third neighbors \cite{Regnault2018}. Table \ref{tab:Tab2} summarizes the six major hopping parameters for $d$-$d$ ($t_i$ ($i=1\ldots,6$)) and two $p$-$d$ ($t_{pd\sigma}$, $t_{pd\pi}$) processes for a nearest-neighbor Co-Co bonding. Following the nearest-neighbor bond and local octahedral coordinate as depicted in Fig.~\ref{fig:Fig4}, among the six hopping parameters $t_1$, $t_3$, $t_4$, $t_5$ denote intraorbital direct hopping between $d_{\rm xz/yz}$, $d_{\rm xy}$, $d_{\rm 3z^2-r^2}$, and $d_{\rm x^2-y^2}$, respectively. The $t_2$ and $t_6$ denote hopping from $d_{\rm xz}$-$d_{\rm yz}$ and $t_{\rm 2g}$-$e_{\rm g}$ mediated by O $p$ orbitals, which include both direct and indirect hoppings. Note that applying the threefold rotation operation generates the other two nearest-neighbor hopping integrals.

Analyzing $d$-$d$ and $p$-$d$ hopping processes for BC$X$O, we make several important observations. Since the in-plane lattice constants increase with the substitution of heavier pnictogen elements at $X$ sites (see Table \ref{tab:Tab1}), Co-Co distance enhances which affects direct $d$-$d$ overlap integrals. From Table \ref{tab:Tab2}, we identify two major hopping channels, namely $t_3$ and $t_6$ as schematically depicted in Fig. \ref{fig:Fig4} that play a crucial role to suppress nearest neighbor Heisenberg exchange. As evident from Table \ref{tab:Tab2}, $t_3$ monotonically decreases while $t_6$ increases simultaneously as a function of pnictogen substitutions. The reduction in $t_3$ hopping integral is attributed to enhancement in Co-Co distance due to the increase in-plane lattice constants with pnictogen substitutions. On the other hand, the enhancement of $t_6$ hopping parameter is ascribed to reduction in Co-O bond distance because of increase in Co-O-Co bond angle from P$\rightarrow$Sb (see Table \ref{tab:Tab1}). For third neighbor, direct hopping strength $t'_5$ is the largest one, which plays an important role to dictate third nearest-neighbor antiferromagnetic Heisenberg exchange.

Next we compute magnetic exchange interactions using hopping parameters from Table \ref{tab:Tab2}. The results are listed in Table \ref{tab:Tab3} for the first- and third-nearest-neighbors. There are total three exchange paths denoted by $A$ ($t_{\rm 2g}$-$t_{\rm 2g}$), $B$ ($t_{\rm 2g}$-$e_{\rm g}$), and $C$ ($e_{\rm g}$-$e_{\rm g}$), respectively. Each exchange process involves three distinct mechanisms, numbered by 1 (direct), 2 (double-hole), and 3 (cyclic) exchange processes, resulting total nine contributions to exchange as described in previous studies \cite{Khaliullin2018,Kee2023}. The direct exchange involves virtual excitations due to inter-site charge fluctuations between two neighboring Co sites ($d^7_id^7_j\rightarrow d^6_id^8_j$). The double-hole process is virtual charge-transfer excitations between two Co $d$ orbitals mediated by an O $p$ ($d^7_ip^6d^7_j-d^8_ip^4d^8_j$), when double-hole are created on the oxygen $p$-orbitals. Lastly cyclic exchange is a quantum mechanical process in which total two holes are created during an intermediate process at two O $p$ orbitals between two neighboring Co sites ($d^7_ip^6_1p^6_2d^7_j\rightarrow d^8_ip^5_1p^5_2d^8_j$). For the nearest-neighbor exchange, all nine different contributions are listed in Table \ref{tab:Tab3}. Only three direct exchange contributions are tabulated for the third-nearest-neighbor terms due to the complexity of enumerating all possible $p$-$d$ virtual processes.

Comparing Heisenberg, Kitaev, and off-diagonal exchanges in Table \ref{tab:Tab3} for the nearest-neighbor part, we find that the ferromagnetic Heisenberg exchange is the most dominant in all BC$X$O. As the pnictogen element changes from P to Sb, the strength of $J_1$ monotonically decreases. Concurrently, antiferromagnetic Kitaev exchange gradually decreases and eventually flips the sign to become ferromagnetic. The values of our estimated exchange interactions for BCAO are consistent with those reported previously, which predicted dominant Heisenberg and weak Kitaev exchange terms \cite{Kee2023}. Furthermore, to unravel how Heisenberg and Kitaev exchange change with pnictogen substitutions, we analyze the contribution of each exchange process to total exchange. Table \ref{tab:Tab3} reveals that $A1$, $B2$, and $B3$ are the three dominant contributions from three different exchange paths in BC$X$O. Because of the opposite sign of $B2$ and $B3$, a large part of exchange interactions from $t_{\rm 2g}$-$e_{\rm g}$ and $e_{\rm g}$-$e_{\rm g}$ paths cancel each other so that total Heisenberg and Kitaev exchange interactions mostly arise from the $t_{\rm 2g}$-$t_{\rm 2g}$ path.

In addition, we find that the largest contribution to nearest-neighbor Heisenberg exchange stems from $t_3$ hopping in BC$X$O, whose strength gradually reduces from P to Sb. On the other hand, the direct exchange process originating from $t_1t_3$ hopping results in largest contribution to nearest-neighbor antiferromagnetic Kitaev exchange for P and As. With the substitution of Sb, $t_6$-derived direct exchange term now dominates over $t_1t_3$ mediated exchange, resulting in the sign reversal of the Kitaev exchange.

Since many cobaltate compounds show significant third-nearest-neighbor exchanges that play a crucial role in geometrical frustration, and also in stabilizing experimentally observed double zigzag magnetic order \cite{Regnault2018}, we compute those from the third-neighbor hopping terms. The results are shown in Table \ref{tab:Tab3}. We comment that due to the overwhelming number of possible oxygen-mediated virtual processes in the third-neighbor exchange interactions, here we only present results from the direct exchange process. We observe finite antiferromagnetic Heisenberg exchange with almost negligible Kitaev interaction. The values of estimated $-J_3/J_1$ ratios are 0.26, 0.43, and 0.22 for BCPO, BCAO, and BCSO, respectively. According to the study of the $J_1$-$J_3$ model, there are three plausible ground states for BC$X$O, namely ferromagnetic for 0<$-J_3/J_1$<0.25, spiral phase for 0.25<$-J_3/J_1$<0.40, and zigzag magnetic ground state beyond 0.40 \cite{Reatto1979,Fouet2001}. Experimentally, $-J_3/J_1$ for BCAO has been reported to be 0.26-0.34 \cite{Regnault2018,Broholm2023} which is indicative of geometrical frustration. Our calculated $-J_3/J_1$ ratios are reasonably comparable to the reported one.

\section*{Discussion}

Based on our first-principles calculations and estimations of magnetic exchange interactions, we comment on a few important aspects of this study. To capture paramagnetic Mott phase and formation active $L_{\rm eff}=1$ orbital degree of freedom, incorporation of dynamical mean-field correlation is essential in BC$X$O. Since the substitution of the pnictogen elements systematically increases the lattice constants, it is effective in controlling the Co-Co direct hopping and hence, the strength of direct exchange interaction. Besides, it can result in switching the sign of Kitaev exchange from antiferromagnetic to ferromagnetic and also, tuning the $-J_3/J_1$ for the realization of the geometric frustration. In addition to those, the size of trigonal crystal-field splitting within $t_{\rm 2g}$ sector in BC$X$O can be tuned via the pnictogen substitution. We predict that, despite the absence of the $X$ = Sb limit of the BC$X$O system, a gradual Sb replacement of As in BCAO via chemical doping can be an interesting way to tune exchange interactions. This may lead to an interesting coexistence of geometric frustration induced by the $J_3$-$J_1$ competition and sizable Kitaev interaction, which might result in a novel phase of matter. 

In summary, our electronic structure calculations of BC$X$O reveal the formation of the paramagnetic Mott phase in all three compounds. Our study reveals the existence of the unquenched effective orbital degree of freedom $L_{\rm eff}=1$ and $S=3/2$ at Co$^{2+}$ ions with sizable trigonal crystal field, which is comparable to the size of atomic spin-orbit coupling of Co. Therefore, with spin-orbit coupling many-body $J_{\rm eff}=1/2$ Kramer's doublet is expected to form in BC$X$O. The analysis of magnetic exchange interactions reveals that in all three compounds, nearest-neighbor Heisenberg exchange is the dominant interaction with finite Kitaev exchange. Finally it is shown that the pnictogen substitution can be a useful tool to tune the exchange interactions and frustrations in the BC$X$O series, highlighting the potential of these compounds in studying novel quantum magnetism.

\section*{Data availability}

The data supporting the findings of this study are available within the manuscript and Supplementary Information. Additional relevant data are available from the corresponding authors upon request.

\section*{Code availability}

The Vienna {\it ab initio} Simulation Package ({\sc vasp}, Ver. 6.3, see https://www.vasp.at) and {\sc wien2k} (Ver. 2019, see http://www.{\sc wien2k}.at) are commercial codes, while DFT+embedded DMFT Functional code (see http://hauleweb.rutgers.edu/tutorials/) is an open-source one which runs on top of the {\sc wien2k} package.

\section*{Methods}

\subsection*{Density functional theory calculations}
DFT calculations were carried out using pseudopotential-based Vienna Ab initio Simulation Package ({\sc vasp}) within projector augmented wave (PAW) formalism \cite{Kresse1996}. The Perdew-Burke-Ernzerhof (PBE) generalized gradient approximation (GGA) adapted for solids \cite{Perdew2008} was chosen as the exchange-correlation functional. Additionally, effective on-site Coulomb potential ($U_{\rm eff}$ = 2, 4 eV) was applied to the $d$ orbitals of Co using a rotationally-invariant DFT+$U_{\rm eff}$ formalism \cite{Dudarev1998}. The planewave cutoff and size of $k$-grids were set to 500 eV and 14$\times$14$\times$14, respectively. The structural optimization was performed using PBEsol+$U$ to get a better agreement of Co-O bond lengths and lattice constants with the experiment. For optimization of crystal structures, force and energy tolerance factors were fixed to be 10$^{-4}$ meV/{\AA} and 10$^{-9}$ eV respectively. It is worth noting that optimizing the crystal structure without the consideration of the on-site Coulomb repulsion and magnetism greatly underestimates Co-O bond length, resulting in spurious metallic behavior in the dynamical mean-field calculations \cite{Kim2022}. Furthermore, to estimate hopping parameters between different Co $d$ orbitals, we employed the maximally-localized Wannier function (MLWF) method \cite{Vanderbilt1997} as implemented in {\sc wannier90} \cite{Pizzi2020} package. Note that the Wannier functions were obtained from DFT calculation without including the $U_{\rm eff}$ and magnetism to avoid double-counting effects of the Coulomb repulsion and exchange splitting.

\subsection*{Dynamical mean-field theory calculations}
First-principles DFT+DMFT calculations were performed within the framework of the local density approximation (LDA) of the Cerpeley-Alder parametrization \cite{Alder1980}, employing the Embedded DMFT Functional ({\sc edmftf}) code \cite{Haule2010,Haule2018}, interfaced with the full-potential {\sc wien2k} \cite{Blaha2018} code. The primitive Brillouin zone of BC$X$O was sampled with a 14$\times$14$\times$14 $k$-grid mesh. The $RK_{\rm max}$ was set to 7.0. The calculation results were obtained using an Ising-type density-density Coulomb interaction and were further verified using the rotationally-invariant full Coulomb interaction. The temperature of the bath was set to be 116 K. The on-site Coulomb repulsion and Hund's coupling parameters were chosen to be ($U, J_{\rm H}$)=(8, 0.8) eV. All calculation results presented below employed a nominal double-counting scheme \cite{Haule2010}. We emphasize that our DMFT results remain qualitatively the same within a range of the values of ($U, J_{\rm H}$) parameters and choices of nominal occupancies \cite{Kim2022}. The single-site quantum impurity problem was solved self-consistently using a hybridization-expansion continuous-time quantum Monte Carlo method, where the Co $d$ orbitals were chosen as the correlated subspace and the number of the total Monte Carlo steps was chosen to be 24$\times$10$^8$. To achieve better accuracy in crystal structures of BC$X$O, DFT+$U_{\rm eff}$-optimized structures were further subject to the structural relaxation within DMFT calculations until the largest Hellmann-Feynman force on each atom reaches below 3 meV/{\AA}.

\begin{acknowledgments}
We thank Hosub Jin, Ara Go, Choong H. Kim, and Kwang-Yong Choi for fruitful discussions. This work was supported by the Korea Research Fellow (KRF) Program and the Basic Science Research Program through the National Research Foundation of Korea funded by the Ministry of Education [Grant No. NRF-2019H1D3A1A01102984, NRF-2020R1C1C1005900, NRF-2022R1I1A1A01071974, RS-2023-00220471]. HSK additionally appreciates the support of the 2021 Research Grant for new faculty members from Kangwon National University.
\end{acknowledgments}

\section*{Author contributions}

This project was conceived by H.-S.K. and S.S. Electronic structure calculations were performed by S.S., H.-S.K., and F.C. All authors discussed the results and contributed to writing the manuscript.

\section*{Competing interests}

The authors declare no competing interests.

\end{document}